\documentclass[preprint,3p]{elsarticle}
\usepackage{dsfont,amsmath}
\biboptions{sort&compress}

\begin{document}
\begin{frontmatter}
\title{Anharmonic oscillator and the optimized basis expansion}
\author{Pouria Pedram}
\ead{p.pedram@srbiau.ac.ir}
\address{Department of Physics, Science and Research Branch, Islamic Azad University, Tehran, Iran}

\begin{abstract}
We introduce various optimization schemes for highly accurate
calculation of the eigenvalues and the eigenfunctions of the
one-dimensional anharmonic oscillators. We present several methods
of analytically fixing the nonlinear variational parameter specified
by the domain of the trigonometric basis functions. We show that the
optimized parameter enables us to determine the energy spectrum to
an arbitrary accuracy. Also, using the harmonic oscillator basis
functions, we indicate that the resulting optimal frequency agrees
with the one obtained by the principle of the minimal sensitivity.
\end{abstract}


\end{frontmatter}

\section{Introduction}\label{sec1}
Eight decades after the discovery of quantum mechanics, the
Schr\"odinger's famous equation still remains an interesting subject
for various investigations, aiming at extending its applications and
at developing more efficient analytic and numerical methods for
obtaining its energy eigenvalues and stationary states. The interest
in this subject ranges from various branches of mathematics,
physics, and chemistry. This has been the driving force behind the
development of perturbative and nonperturbative methods for this
kind of problems. Among them are the factorization method
\cite{Infeld,pedramAJP}, semiclassical approximation
\cite{slavyanov}, finite-difference technique \cite{Witwit},
optimized Rayleigh-Ritz variational scheme \cite{Okopinska,Koscik},
variational matrix solution \cite{Balsa}, instanton method
\cite{Kleinert}, transfer matrix method \cite{Zhou} and many other
specific methods.

One general approach to construct a continuous wave function
$\psi(x)$ is to represent its values on a set of mesh (lattice)
points $x_n$ which is the starting point of the various mesh
methods. Although this approach is simple, it is usually very
inaccurate which is due to the fact that it only contains the local
information of the wave function. To overcome this problem, Schwartz
proposed a method based on a global construction of an approximate
wave function which involves only the values $\psi_n$ at the
selected mesh points \cite{Schwartz}. He showed that using an
optimized mesh spacing the obtained errors are as small as $A^{-N}$
or even $1/N!$ where $N$ is the number of the mesh points. This
shows the priority of this method over the usual numerical methods
that yield errors as small as $1/N$, $1/N^2$, etc. The application
of a set of orthogonal functions on a finite domain for solving a
wide class of problems such as function interpolation and the
numerical solution of the Schr\"odinger equation has attracted much
attention in recent years (see Ref.~\cite{Amore} and references
therein). For instance, the variational sinc collocation based on
the principle of minimal sensitivity (PMS) method can be effectively
used to obtain the highest precision with a given number of mesh
points \cite{Amore2}.

The second popular scheme is to expand the wave function in terms of
an orthonormal set of the eigenfunctions of a Hermitian operator,
i.e., the basis-set expansion method. For instance, the
trigonometric basis functions obeying Dirichlet boundary condition
(particle in a box basis) can be effectively used to find the
spectrum of an unbounded problem. The low lying energy levels are
approximately equal to the exact ones with high accuracy, if the
boundedness parameter is in near vicinity of an optimal value upon
implementing the Rayleigh-Ritz variational method
\cite{taseli,Bhattacharyya}. The extension of this method for the
periodic boundary condition is also discussed in
Ref.~\cite{pedramMolPhys}. For these cases, the variational
parameter is the width of the finite interval (the size of the box).
On the other hand, if we expand the wave function in terms of the
harmonic oscillator eigenfunctions, the frequency of the oscillator
would be the variational parameter.

In this paper, first we briefly outline the Schwartz's method and
its error analysis which is based on the large distance behavior of
the wave function. Then by defining the Schwartz's length, which as
we shall show has a useful application in diagonalization of the
Hamiltonian with the particle in a box basis set, the nearly
accurate results can be obtained if we use this length as the
optimal width of the finite domain. By imposing a physically
acceptable relation between the potential energy at the optimal
length and the maximum available basis eigenenergy, we explain the
physics behind the Schwartz's length and obtain the optimal length
without need to the error analysis which is done in the Schwartz's
original paper. As it is emphasized in Ref.~\cite{Schwartz}, there
is a close connection between the Schwartz's scheme and the Fourier
expansion where we shall elaborate it in the next section. In this
view, we improve our estimation and introduce some alternative and
more accurate optimal lengths for the particle in a box basis
functions. Also we present the optimal frequency for the expansion
of the wave function in terms of the harmonic oscillator basis
functions.

Another way for finding the optimal length is using the stationarity
of the trace of the Hamiltonian \cite{Okopinska,Koscik}. This method
is based on the principle of minimal sensitivity but demands on
fixing the values of nonlinear parameters before diagonalization of
the truncated matrix. In fact, this optimal value extremizes the
trace and results in highly accurate results. Here we apply this
formalism for the trigonometric basis functions and find the related
optimal length for the anharmonic oscillators.

\section{The Schwartz's method}\label{sec2}
Let us consider an analytic reference function $u(x)$ which has
simple zeros at the real points $x=x_n$ to approximate the wave
function $\psi(x)$. We can define an interpolating wave function
$\overline{\psi}(x)$ to approximate $\psi(x)$ as
\begin{equation}\label{psi}
\overline{\psi}(x)=\sum_m \psi_m
\frac{u(x)}{x-x_m}\frac{1}{u'(x_m)},
\end{equation}
where $\psi_m=\psi(x_m)$. So at the mesh points $x=x_n$, the
interpolating wave function takes the same values as $\psi(x)$
there. With this definition, the derivatives of $\overline{\psi}$ at
the mesh points
$\displaystyle\frac{d\overline{\psi}}{dx}\bigg|_{x_n}$ or its
integrals $\displaystyle\int_{x_0}^{x_n}\overline{\psi}(x)\mathrm{d}x$ only
depend on $\psi_m$ not the derivatives or integrals (see
Ref.~\cite{Schwartz} for details). However, all $\psi_m$ would
contribute to construct the corresponding values of
$\overline{\psi}$ which means that the global information about the
wave function $\psi$ is used to find the approximation.

We can write the above equation as an exact relation by introducing
the error term $\epsilon$ as
\begin{equation}
\psi(x)=\sum_m \psi(x_m)
\frac{u(x)}{x-x_m}\frac{1}{u'(x_m)}+\epsilon.
\end{equation}
Note that, we are usually interested for the cases where the wave
functions decreases rapidly for large $x$ (bound state solutions),
so effectively the infinite sum over the mesh points can be
truncated to a finite sum. In fact, there are two sources of error
in our analysis. First one $\epsilon_A$ due to the analytical
approximation and the second $\epsilon_T$ due to the truncation. If
we make $\epsilon_A$ approximately equal to $\epsilon_T$ by choosing
a relation between the mesh spacing $h$ and the truncation at $n<N$,
the total error will be reduces considerably. This will prevent us
from using a too small $h$ when the truncation error dominates or
using a too large cutoff when the mesh error has the dominant role.

To compute the errors, consider a bound state wave function which
has the following behavior at large distances
\begin{equation}\label{psi-large}
\psi(x)\sim e^{-ax^p},\hspace{2cm}\mbox{for large}\, x,
\end{equation}
so the truncation error reads
\begin{equation}\label{error1}
\epsilon_T\approx e^{-a(Nh)^p}.
\end{equation}
For the mesh size error, we need to perform a contour integral in
the complex plane. By taking the reference function as
$u(x)=\sin(\pi z/h)$, the integral can be estimated by the
stationary phase method and we obtain \cite{Schwartz}
\begin{equation}\label{error2}
\epsilon_A\approx e^{-bh^{-q}},
\end{equation}
where $q=p/(p-1)$ and
\begin{equation}
b=\left(\frac{\pi^p}{ap}\right)^{1/(p-1)}\left(\frac{p-1}{p}\right)\sin\left[\frac{\pi}{2(p-1)}\right].
\end{equation}
Now the optimal value of $h$ for each $N$ can be obtained by
equating Eqs.~(\ref{error1}) and (\ref{error2})
\begin{equation}
h_{S}=\left(\frac{b}{aN^p}\right)^{\frac{p-1}{p^2}},
\end{equation}
which results in the exponential decrease of the error by increasing
the number of the mesh points
\begin{equation}
\epsilon\approx e^{-CN},
\end{equation}
where $C=b(a/b)^{1/p}$. To apply the method, let us consider the
following dimensionless time-independent one-dimensional
Schr\"{o}dinger equation\footnote{Note that for $V(x)=\beta x^k$ we
have $E\rightarrow \beta^{\frac{2}{k+2}}E$.}
\begin{equation}\label{Schrodinger}
\left(-\frac{d^2}{dx^2}+x^k\right)\psi(x)=E\,\psi(x),\hspace{2cm}k=2,4,6,\ldots.
\end{equation}
For this case, the wave function has the asymptotic behavior for
large $x$ given by (\ref{psi-large}) with
\begin{equation}
p=\frac{k+2}{2},\hspace{2cm}a=\frac{2}{k+2},
\end{equation}
which results in
\begin{equation}
b=\pi^{\frac{k+2}{k}}\left(\frac{k}{k+2}\right)\sin\left(\frac{\pi}{k}\right),\hspace{.5cm}\mathrm{and}\hspace{.5cm}h_{S}=\left[
\frac{1}{2}k\,\pi^{\frac{k+2}{k}}\sin\left(\frac{\pi}{k}\right)\right]^{\frac{2k}{(k+2)^2}}N^{-\frac{k}{k+2}}.
\end{equation}
Now if we define the Schwartz's length $L_S\equiv Nh_S$, we have
\begin{equation}\label{LS}
L_{S}(N)=\left[
\frac{1}{2}k\,\pi^{\frac{k+2}{k}}\sin\left(\frac{\pi}{k}\right)\right]^{\frac{2k}{(k+2)^2}}N^{\frac{2}{k+2}},
\end{equation}
where, as we shall show in the next section, it can be used as an
accurate candidate for the optimal length in the context of the
Fourier expansion of the wave function.

At this point, it is worth to mention the connection between the
above collocation method and the Fourier expansion scheme. So let us
define the generalized sinc functions as
\begin{equation}
S_m(h,x) \equiv \frac{\sin\left[\pi (x-m h)/h\right]}{\pi (x-m
h)/h}, \label{sinc}
\end{equation}
where $m \in\mathds{Z}$, uniform grid spacing $h$ and
$x\in\mathds{R}$. Now using Eq.~(\ref{psi}) and $u(x)=\sin(\pi x/h)$
we obtain
\begin{equation}
\overline{\psi}(x)=\sum_m \psi_m S_m(h,x),
\end{equation}
which defines the sinc collocation method. It is also possible to
write a similar equation in terms of the little sinc functions
\cite{Amore}. Consider an orthonormal set of particle in a box basis
functions vanishing at $x=\pm L$
\begin{equation}
\varphi_n(x) = \frac{1}{\sqrt{L}} \sin \left[\frac{n\pi}{2L}
(x+L)\right],\hspace{2cm}n=1,2,\ldots,
\end{equation}
and define
\begin{eqnarray}
\overline{\delta}_N(x,y) &=& \frac{2L}{N} \sum_{n=1}^N \varphi_n(x)
\varphi_n(y), \nonumber\\
&=& \frac{1}{2N} \ \left\{ \frac{\sin\left[\frac{(2 N+1) \pi
(x-y)}{4 L}\right]}{\sin\left[\frac{\pi (x-y)}{4 L}\right]}
-(-1)^{N}  \frac{\cos\left[\frac{(2 N+1) \pi (x+y)}{4 L}\right]}{
\cos\left[\frac{\pi(x+y)}{4 L}\right]} \right\},
\end{eqnarray}
where $N$ takes even values. Because of the completeness of the basis functions we have
\begin{equation}
\lim_{N\rightarrow \infty} \frac{N}{2L}\overline{\delta}_N(x,y) =
\delta(x-y).
\end{equation}
By setting $h =2L/N$, $y_k = k h$ and selecting even values of $N$,
we define the set of $(N-1)$ little sinc functions (LSF) as
\begin{equation}
s_k(h,N,x) \equiv  \frac{1}{2 N} \left\{ \frac{\sin\left[
\left(1+\frac{1}{2N}\right) \ \frac{\pi}{h} (x-k h)\right]}{
\sin\left[\frac{\pi}{2 N h} (x-k h)\right] }
-\frac{\cos\left[\left(1+\frac{1}{2N}\right) \ \frac{\pi}{h} (x+k
h)\right]}{ \cos\left[\frac{\pi}{2 N h} (x+k h)\right]} \right\}.
\label{sincls}
\end{equation}
Therefore, LSF become the standard sinc functions when $N$ goes to
infinity, i.e.,
\begin{equation}
\lim_{N\rightarrow \infty} s_k(h,N,x) = \frac{\sin[\pi (x-k
h)/h]}{\pi (x-k h)/h}= S_k(h,x).
\end{equation}
The LSF have some common properties with the sinc functions, for
instance, we can approximate the wave function on the interval
$(-L,L)$ as
\begin{equation}
\overline{\psi}(x)=\sum_m \psi_m s_m(h,N,x),\label{23}
\end{equation}
where can be understood using the definition of $s_k(h,N, x)$ in
terms of the completeness relation. Therefore, we can rewrite
Eq.~(\ref{23}) as
\begin{equation}
\overline{\psi}(x)=\sum_m \left[h\sum_k
\psi_k\varphi_m(x_k)\right]\varphi_m(x).
\end{equation}
In the limit $N\rightarrow\infty$ this relation becomes
\begin{equation}
\psi(x)=\sum_m \left[\int_{-L}^{L}
\psi(x)\varphi_m(x)\mathrm{d}x\right]\varphi_m(x)=\sum_m a_m \varphi_m(x),
\end{equation}
which is the well-known Fourier expansion. So there is a close
relation between the sinc collocation method and the trigonometric
basis expansion.

\section{The trigonometric basis expansion}\label{sec3}
In this section we study the diagonalization of the Hamiltonian in
terms of the particle in a box eigenfunctions that is basically
different from the Schwartz's method. Then by imposing a constraint
on the potential energy at the optimal length and the maximum
available basis eigenenergy, we analytically fix the variational
parameter. Before going further note that in the Schwartz's scheme
$N$ is the number of mesh points whereas in this section $N$ is the
number of basis functions. However, in the Schwartz's method the
reference function vanishes at the $N$ mesh points and the $N$
trigonometric basis functions have at most $N$ nodes. Also we have
$Nh=L$. So we expect that there would be a close connection between
the Schwartz's method and the trigonometric basis expansion which is
also explicitly elaborated in the previous section.

For the potentials which are even functions of $x$, to avoid large
matrices, we can use
\begin{equation}
\phi_m(x)=\sqrt{\frac{1}{L}}\cos\left[\left(m-\frac{1}{2}\right)\frac{
\pi
x}{L}\right],\hspace{.5cm}\mathrm{and}\hspace{.5cm}\phi_m(x)=\sqrt{\frac{1}{L}}\sin\left(\frac{m
\pi x}{L}\right),
\end{equation}
basis functions ($m=1,2,\ldots,N$) for even and odd parity
solutions, respectively, and write the wave function as $\psi(x)=
\sum_{m=1}^N A_{m} \phi_m(x)$ which vanishes at $\pm L$. Now the
approximate solutions are the eigenvalues and the eigenfunctions of
the $(N\times N)$ Hamiltonian matrix $\mathbf{H}_N$ where can be
written as
\begin{equation}\label{even}
H_{mn}=\displaystyle\left(m-\frac{1}{2}\right)^2\frac{\pi^2}{L^2}\delta_{mn}
+\left(\frac{L}{\pi}\right)^{k} \bigg(D_{m+n-1}+D_{m-n}\bigg),
\end{equation}
and
\begin{equation}\label{odd}
H_{mn}=\displaystyle\frac{m^2\pi^2}{L^2}\delta_{mn}
+\left(\frac{L}{\pi}\right)^{k} \bigg(D_{m-n}-D_{m+n}\bigg),
\end{equation}
for even and odd states, respectively. Here, $\delta_{mn}$ is the
kronecker's delta and $D_s$ is defined as
\begin{eqnarray}
D_s=\frac{1}{\pi}\int_0^{\pi}\mathrm{d}x\,x^{k}\cos(sx)= \left\{
\begin{array}{ll}
\displaystyle\sum_{i=0}^{\frac{k}{2}-1}\frac{(-1)^{i+s}}{s^{2(i+1)}}\frac{k!}{(k-2i-1)!}\pi^{k-2i-2},&\hspace{.5cm}s>0,\\
\displaystyle\frac{\pi^{k}}{k+1},&\hspace{.5cm}s=0.
\end{array}\right.
\end{eqnarray}
Now the good strategy is to choose a relation between the basis
domain $L$ and the truncation at $n<N$ so that the errors due to the
basis domain and the truncation approximately suppress each other.
This will prevent us from choosing a too large domain for small $N$
or a too large cutoff for small $L$.

\begin{figure}
\begin{center}
\includegraphics[width=8cm]{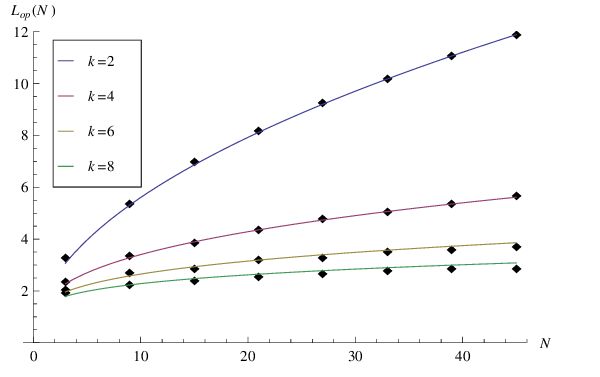}
\caption{\label{fig1}The variationally obtained optimal lengths
(black diamonds) versus $N$ for $V(x)=x^k$ and the predicted optimal
length curves (\ref{Lop}).}
\end{center}
\end{figure}

Note that, the expansion the solutions in terms of particle in a box
basis functions approximately corresponds to confining the potential
in an infinite potential well,\footnote{This correspondence is exact
as $N$ goes to infinity.} i.e., $V(x)=x^{k}$ for $|x|<L$ and
$V(x)=\infty$ elsewhere. Moreover, each energy eigenvalue is a
superposition in the form $E_n=\sum_{m=1}^{N}P_{nm}\varepsilon_m$
where $\sum_{m=1}^{N}P_{nm}=1$, $P_{nm}\geq0$, and
$\varepsilon_n=n^2\pi^2/L^2$. Since for $|x|\ge L$ this model is not
identical with the original potential, the basis functions with
energies larger than $V(L)$ would not have a useful contribution to
the sought-after solutions. In this case, we lose accuracy which is
due to the small well's width. Also, when $V(L)\gg \varepsilon_N$
the solutions would be inaccurate which means that $L$ is too large.
Keeping these two points in mind, we conclude that the value of the
potential at $L_{\mathrm{op}}$ should be proportional to the maximum
energy of the basis functions $\varepsilon_N$, namely
\begin{eqnarray}\label{VL}
L_{\mathrm{op}}^k=\alpha(k)\frac{N^2\pi^2}{L_{\mathrm{op}}^2},
\end{eqnarray}
where $\alpha(k)\geq1$ and it is of order of one. From this equation
we can find the optimal value of $L$ as
\begin{eqnarray}\label{VL2}
L_{\mathrm{op}}(N)=\bigg(\pi^2\alpha(k)\bigg)^{\frac{1}{k+2}}N^{\frac{2}{k+2}},
\end{eqnarray}
which has the same functional form as $L_S(N)$ (\ref{LS}). This
similarity is due to the relation $L=Nh$ and the fact that the
maximum number of nodes in this method agrees with the number of
mesh points in the Schwartz's method which both are represented by
$N$. Now, it is only remained to determine $\alpha(k)$ which can be
found by choosing an acceptable ansatz and comparing with the
variationally obtained results (Fig.~\ref{fig1}).\footnote{From now
on we show the relative error of the energy spectrum by
$\epsilon_n\equiv\left|\frac{E_n-E_n^{\mathrm{exact}}}{E_n^{\mathrm{exact}}}\right|$.}
It is straightforward to check that for the simple harmonic
oscillator ($k=2$), the most accurate solutions can be obtained when
the value of the potential at $L_{\mathrm{op}}$ is equal to the
maximum energy of the basis functions $\varepsilon_N$, i.e.,
$\alpha(2)=1$. For this case using (\ref{VL2}) we have
$L_{\mathrm{op}}=\sqrt{\pi N}=L_S$. For other values of $k$, the
simplest choice is looking for a relation in the form
$\alpha(k)=\eta^{\frac{k-2}{2}}$ with constant $\eta$ to ensure
$\alpha(2)=1$. As Fig.~\ref{fig1} shows, we can properly fit
Eq.~(\ref{VL2}) to the variational values upon choosing $\eta=\pi/2$
which results in
\begin{eqnarray}\label{Lop}
\alpha_{\mathrm{op}}(k)=\left(\frac{\pi}{2}\right)^{\frac{k-2}{2}},\hspace{1cm}\mathrm{and}\hspace{1cm}L_{\mathrm{op}}(N)=\sqrt{\frac{\pi}{2^{\frac{k-2}{k+2}}}}\,N^{\frac{2}{k+2}},
\end{eqnarray}
Note that, although the coefficients of $N^{\frac{2}{k+2}}$ in $L_S$
(\ref{LS}) and $L_{\mathrm{op}}$ (\ref{Lop}) seems to be very
different, they are approximately equal especially for $k<10$. For
instance, for $k=4$ we have
$L_{\mathrm{op}}=\sqrt{\pi/2^{1/3}}N^{1/3}\simeq1.579N^{1/3}$ which
is nearly equal to $L_S=2^{1/9}\pi^{1/3}N^{1/3}\simeq1.582N^{1/3}$.
In Table \ref{tab}, we have reported the accuracy of the energy
eigenvalues for $k=\{2,4,6,8\}$ which are in complete agreement with
the variationally obtained solutions \cite{taseli}.

\begin{figure}
\centerline{\begin{tabular}{ccc}
\includegraphics[width=8cm]{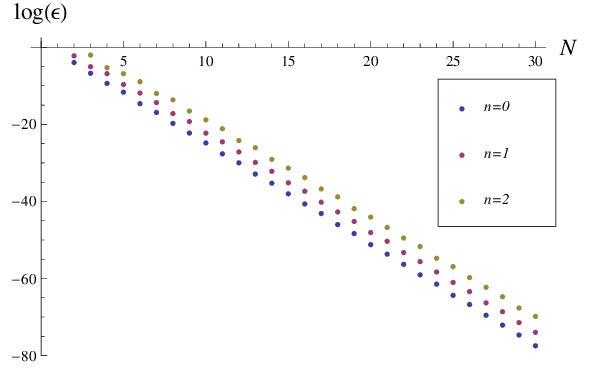}
 &\hspace{0.cm}&
\includegraphics[width=8cm]{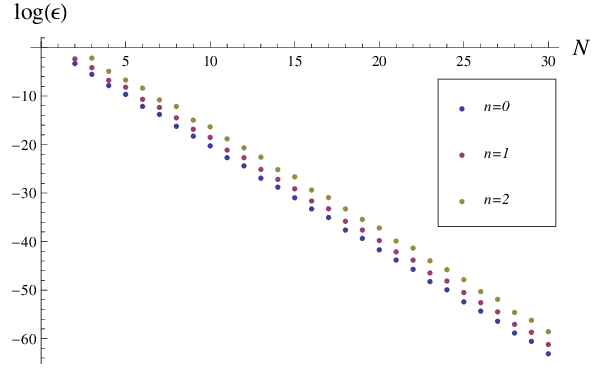}
\end{tabular}}
\caption{\label{figk}The error versus $N$ for $k=4$ and
$L_{\mathrm{op}}$ (left), and for $k=6$ and $L_S$ (right).}
\end{figure}

Since $\alpha(k)$ for $k\gg1$ will not be of ${\cal O}(1)$, we
expect that the validity of Eq.~(\ref{Lop}) breaks down for large
$k$. At this limit we have $V(L_{\mathrm{op}})\gg \varepsilon_N$
which would result in inaccurate solutions as a consequence of too
large $L_{\mathrm{op}}$. To check this point, we consider the
problem of $V(x)= x^{k}$ where $k\rightarrow\infty$. For this case,
the predicted optimal length is
$\lim_{k\rightarrow\infty}L_{\mathrm{op}}(N)\simeq
\sqrt{\pi/2}\simeq1.253$. However, the optimal length predicted by
the Schwartz's formula has the correct limiting value, i.e.,
$\lim_{k\rightarrow\infty}L_S(N)=1$ and gives the following
proportionality coefficient:
\begin{eqnarray}\label{alphaS}
\alpha_S(k)=\left[\frac{k}{2}\sin\left(\frac{\pi}{k}\right)\right]^{\frac{2k}{k+2}},
\end{eqnarray}
where $\alpha_S(2)=1$ and as we have desired, it is of the order of
unity for all $k$, i.e., $1\leq \alpha_S\leq\pi^2/4$. Also,
Fig.~\ref{figk} shows that the error exponentially decreases with
respect to the number of the basis functions. In comparison, the
calculations show that the usage of $L_S$ results in more accurate
energy spectrum than those obtained by $L_{\mathrm{op}}$.

\begin{table}
\begin{center}
\begin{tabular}{ccccc}\hline
 $k$&$L_{\mathrm{op}}$   &$N$    &$n$    &$\epsilon_n$
 \\\hline\\
 2  &$\sqrt{30\pi}$     &$30$   &0      & 2.77$\times 10^{-40}$   \\
    &           &       &2      & 9.57$\times 10^{-37}$     \\
    &           &       &4      & 1.47$\times 10^{-33}$     \\
 4  & $\displaystyle\frac{\sqrt{\pi}\,35^{1/3}}{2^{1/6}}$     &35     &0      & 4.95$\times 10^{-40}$       \\
    &           &       &2      & 1.72$\times 10^{-38}$       \\
    &           &       &4      & 1.04$\times 10^{-36}$   \\
 6  & $\sqrt{2\pi}\,5^{1/4}$    &40     &0    &3.31$\times 10^{-36}$       \\
    &           &       &2      & 1.57$\times 10^{-35}$        \\
    &           &       &4      & 8.22$\times 10^{-35}$         \\
 8  & $\displaystyle\frac{\sqrt{\pi}\,45^{1/5}}{2^{3/10}}$     &45     &0      & 2.03$\times 10^{-32}$       \\
    &           &       &2      & 7.19$\times 10^{-32}$         \\
    &           &       &4      & 4.57$\times 10^{-31}$
    \\\hline
\end{tabular}
\end{center} \caption{The relative errors of the energy spectrum of the anharmonic oscillator
$V(x)=x^{k}$.}\label{tab}
\end{table}

\section{Other applications}
The applicability of the introduced optimal lengths is not
restricted to the particular form of the potentials, boundary
conditions, or differential equations. Indeed, the optimal length
can be effectively used for the following issues:

\subsection{Polynomial potentials}
Let us consider a symmetric polynomial potential
$V(x)=\sum_{i=2}^{k}a_{i}x^{i}$ where $a_{k}>0$ and
$i=2,4,\ldots\,$. Since the optimal length related to each term
behaves in a separate manner, we cannot simply use Eq.~(\ref{Lop})
for this polynomial potential. However, for the large values of
$L_{\mathrm{op}}$, the dominant term near the boundaries is $x^{k}$.
In other words, we can also use this equation for the polynomial
potentials when we work with a large set of the basis functions. To
elaborate this fact, let us study the doubly anharmonic oscillator
$V(x)=a_{2}x^{2}+a_{4}x^{4}+a_{6}x^{6}$. The ground state wave
function of this potential should not have nodes and it should
vanish as $x\rightarrow\pm\infty$. Thus, we can examine the
following solution
\begin{eqnarray}
\Psi_0(x)=\exp\left(-\frac{1}{4}b_4x^4+\frac{1}{2}b_2x^2\right),
\hspace{1cm} b_4>0.
\end{eqnarray}
It is easy the check that this trial wave function satisfies the
corresponding Schr\"odinger with the eigenvalue $E_0=-b_2$,
$b_2=\displaystyle-\frac{1}{2}a_4a_6^{-1/2}$, $b_4=a_6^{1/2}$ and a
constraint on the potential \cite{taseli}
\begin{eqnarray}
a_2=b_2^2-3b_4.
\end{eqnarray}
For instance, the potential $V(x)=-2x^{2}+2x^{4}+x^{6}$ obeys this
constraint and have the following ground state eigenvalue and
eigenfunction
\begin{eqnarray}
\Psi_0(x)=\exp\left(-\frac{1}{4}x^4-\frac{1}{2}x^2\right),
\hspace{1cm} E_0=1.
\end{eqnarray}
For this case, using the optimal length related to $k=6$, i.e.,
$L_{\mathrm{op}}=\left(\pi^2N/2\right)^{1/4}$, one can easily obtain
the ground state energy $E_0$ with $28$ and $36$ significant digits
accuracy for $N=30$ and $N=40$ basis functions, respectively.

\subsection{Periodic boundary condition}
We have deduced Eq.~(\ref{Lop}) from diagonalization scheme for the
Hamiltonian with the basis functions which are vanishing at the
boundaries, but we can also use it for the case of the periodic
boundary condition under some circumstances. Note that, for this
case, we observe an inflection point instead of a minimum value in
the graph of the energy versus the domain of basis functions
\cite{pedramMolPhys}. However, since for the large optimal lengths
the low-lying wave functions are almost zero at the boundaries,
Eq.~(\ref{Lop}) is still valid for case of the periodic boundary
condition when $L_{\mathrm{op}}$ is large enough.

\subsection{Multidimensional problems}
In a $d$-dimensional space which the Hamiltonian is invariant under
parity transformation, the basis functions are in the form
\begin{eqnarray}
\phi_{m_1m_2\cdots
m_d}^{\mathrm{even}}(\vec{x})&=&\prod_{i=1}^{d}\sqrt{\frac{1}{L_i}}\cos\left[\left(m_i-\frac{1}{2}\right)\frac{
\pi x_i}{L_i}\right],\\
\phi_{m_1m_2\cdots
m_d}^{\mathrm{odd}}(\vec{x})&=&\prod_{i=1}^{d}\sqrt{\frac{1}{L_i}}\sin\left(\frac{
m_i\pi x_i}{L_i}\right),
\end{eqnarray}
where $m_i=1,2,\ldots,N$. Now the Hamiltonian is expressed as a
$N^d\times N^d$ matrix. Therefore, even for the small values of $N$,
the resulting matrix is too large and the ordinary variational
scheme cannot be applied efficiently. In general, we need to find
$d$ optimal lengths for the accurate calculation of the eigenvalues
and the eigenfunctions. For instance, for the separable potential
$V(\vec{x})=\sum_{i=1}^d V(x_i)$, we can use the introduced optimal
length (\ref{Lop}) which needs to be properly chosen for each
direction to get the desired accuracy. Moreover, if the Hamiltonian
has the rotational symmetry, i.e., $V(\vec{x})=V(|\vec{x}|)$, we
only need one optimal length to diagonalize the $N\times N$
Hamiltonian.

\subsection{Wheeler-DeWitt equation}
One approach to quantize gravity is based on the well-known
Wheeler-DeWitt equation (WDW), i.e., $\mathcal{H}\Psi=0$ where
$\Psi$ is the wave function of the universe \cite{DeWitt}. After
freezing out many degrees of freedom, WDW equation is expressed in
the minisuperspace and $\Psi$ is a function of just a few variables
such as the scale factor, the scalar field, etc. This equation is a
hyperbolic partial differential equation and can be cast in the form
of the Schr\"odinger-like equation in the vicinity of the
boundaries. To solve the WDW equation, the proper adjusting the
width of the domain of the basis functions with the number of the
basis functions is also an important issue which results in accurate
solutions \cite{p1,p2,p3}.

\section{Alternative proposals}\label{sec4}
\begin{figure}
\centering
\includegraphics[width=8cm]{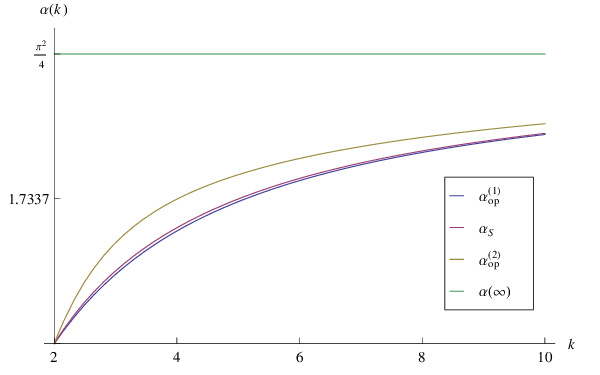}
\caption{\label{figa} The proportionality coefficient $\alpha(k)$.}
\end{figure}

To improve the estimation for the optimal length, we need to modify
Eq.~(\ref{Lop}) in such a way that it takes a finite value as $k$
goes to infinity. For instance, consider the following proposal
\begin{eqnarray}\label{Lop2}
\alpha_{\mathrm{op}}^{(1)}(k)=\left(\frac{\pi}{2}\right)^{\frac{k-2}{k/2}},
\hspace{1cm}\mathrm{and}\hspace{1cm}
L_{\mathrm{op}}^{(1)}(N)=\left(\frac{\pi^{4(k-1)}}{4^{k-2}}\right)^{\frac{1}{k(k+2)}}\,N^{\frac{2}{k+2}}.
\end{eqnarray}
For $k=2$ and $k=4$ this proposal coincides  with Eq.~(\ref{Lop})
and for large $k$ it agrees with Eq.~(\ref{alphaS}), i.e.,
\begin{eqnarray}
\lim_{k\rightarrow\infty}\alpha_{\mathrm{op}}^{(1)}(k)=\lim_{k\rightarrow\infty}\alpha_S(k)=\frac{\pi^2}{4}.
\end{eqnarray}
Since $\alpha_{\mathrm{op}}^{(1)}$ is nearly equal to $\alpha_S$ for
all $k$ (see Fig.~\ref{figa}), the accuracy of this proposal is of
the order of $\alpha_S$.

Up to now, the most accurate proposal for the optimal length is
$L_S$. However, the calculations show that to find more accurate
results, we need to slightly increase $L_S$ for all $k$ and $N$.
Therefore, since we always have $L_{\mathrm{op}}^{(1)}<L_S$, the
accuracy of Eq.~(\ref{Lop2}) is slightly lesser than Eq.~(\ref{Lop})
which can be also confirmed by explicit calculations. To find an
optimal length that is more accurate than $L_S$, note that the
factor $\pi/2$ in Eq.~(\ref{Lop2}) is the first term in the
asymptotic expansion of the term
$\displaystyle\frac{k}{2}\sin\left(\frac{\pi}{k}\right)$ in
Eq.~(\ref{alphaS}), i.e.,
\begin{eqnarray}
\frac{k}{2}\sin\left(\frac{\pi}{k}\right)=\frac{\pi}{2}-\frac{\pi^3}{12k^2}+\mathcal{O}(k^{-4}).
\end{eqnarray}
So a good idea is to add the second factor $\frac{\pi^3}{12k^2}$ to
$\pi/2$ in Eq.~(\ref{Lop2}) and write the optimal length as
\begin{eqnarray}\label{Lop3}
\alpha_{\mathrm{op}}^{(2)}(k)=\left(\frac{\pi}{2}+\frac{\pi^3}{12k^2}\right)^{\frac{k-2}{k/2}}=
\left(1+\frac{\pi^2}{6k^2}\right)^{\frac{k-2}{k/2}}\alpha_{\mathrm{op}}^{(1)}(k),
\end{eqnarray}
and
\begin{eqnarray}
L_{\mathrm{op}}^{(2)}(N)=\left(\frac{\pi^{4(k-1)}}{4^{k-2}}\right)^{\frac{1}{k(k+2)}}
\left(1+\frac{\pi^2}{6k^2}\right)^{\frac{2(k-2)}{k(k+2)}}\,N^{\frac{2}{k+2}}=
\left(1+\frac{\pi^2}{6k^2}\right)^{\frac{2(k-2)}{k(k+2)}}L_{\mathrm{op}}^{(1)}(N),
\end{eqnarray}
which satisfies $\alpha_{\mathrm{op}}^{(2)}(2)=1$ and
$\lim_{k\rightarrow\infty}\alpha_{\mathrm{op}}^{(2)}(k)=\pi^2/4$. It
is straightforward to check that $L_{\mathrm{op}}^{(2)}$
($\alpha_{\mathrm{op}}^{(2)}$) is always slightly greater than $L_S$
($\alpha_S$) (see Fig.~\ref{figa}). Indeed, calculations show that
between the several mentioned candidates for the optimal length,
$L_{\mathrm{op}}^{(2)}$ gives the most accurate energy spectrum. As
a simple application, in Table \ref{tabG}, we have reported the
ground state energy of the anharmonic oscillators using only one
basis function ($N=1$) and the optimal length
$L_{\mathrm{op}}^{(2)}$. For these cases, the matrix of the
Hamiltonian has only one element and the relative error is less than
$20\%$.

\begin{table}
\begin{center}
\begin{tabular}{cccc}\hline
 $k$& $E_0$  &$E_0^{\mathrm{exact}}$ & $\epsilon_0$\\\hline\vspace{.2cm}
2   & 1.19  &   1.00   &  1.9$\times 10^{-1}$     \\\vspace{.2cm} 4
& 1.23  &   1.06   & 1.6$\times 10^{-1}$     \\\vspace{.2cm}
6   & 1.34  &   1.14   &  1.7$\times 10^{-1}$     \\
8   & 1.45  &   1.23   &  1.8$\times 10^{-1}$    \\\hline
\end{tabular}
\end{center} \caption{The ground state energies and their relative errors using one basis function ($N=1$) and $L_{\mathrm{op}}^{(2)}$.}\label{tabG}
\end{table}

\section{Stationarity of the trace of the Hamiltonian}\label{sec5}
\begin{figure}
\centerline{\begin{tabular}{ccc}
\includegraphics[width=8cm]{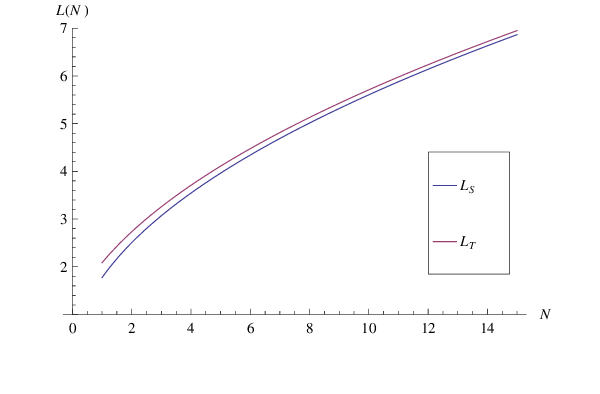}
 &\hspace{0.cm}&
\includegraphics[width=8cm]{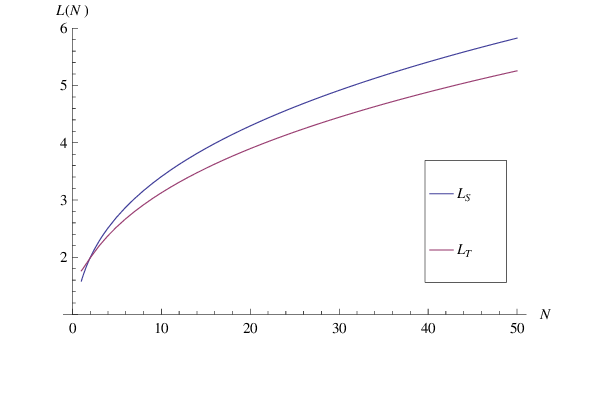}
\end{tabular}}
\caption{\label{fig2}The optimal lengths versus $N$ for $k=2$ (left)
and $k=4$ (right).
}
\end{figure}

\begin{figure}
\centerline{\begin{tabular}{ccc}
\includegraphics[width=8cm]{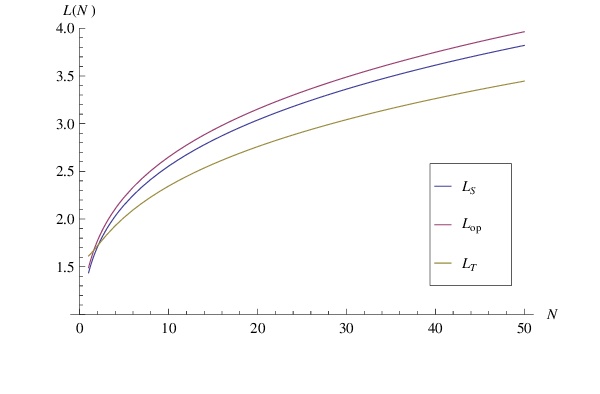}
 &\hspace{0.cm}&
\includegraphics[width=8cm]{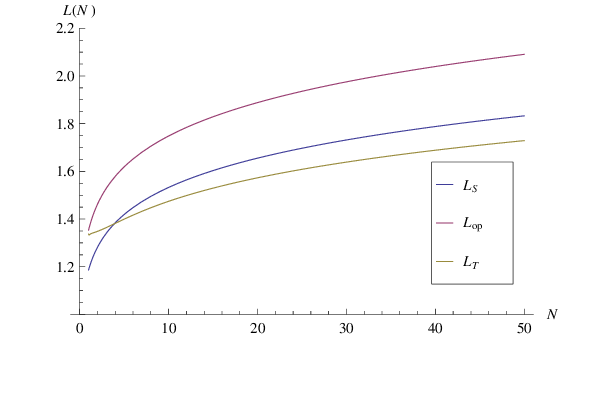}
\end{tabular}}
\caption{\label{fig3}The optimal lengths versus $N$ for $k=6$ (left)
and $k=16$ (right).}
\end{figure}

\begin{figure}
\centering
\includegraphics[width=8cm]{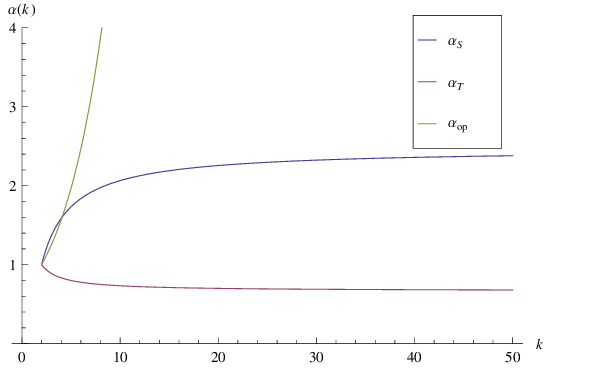}
\caption{\label{fig5} The proportionality coefficient versus $k$.}
\end{figure}

In this section we apply the optimized Rayleigh-Ritz method proposed
by Okopinska \cite{Okopinska,Koscik} for the trigonometric basis
functions. This method is originally used for the harmonic
oscillator eigenfunctions and it is based on fixing the values of
nonlinear parameters before diagonalization of the truncated matrix.
Before diagonalization, the only physical quantity that can be
determined is the trace of the Hamiltonian
\begin{eqnarray}
\mathrm{Tr}_NH=\sum_{n=1}^{N}\langle n|\hat{H}|n\rangle,
\end{eqnarray}
which represents the $N$th-order approximation to the sum of
energies of the $N$ lowest bound states. Now the strategy is to
choose the the values of nonlinear parameters so as to make
$\mathrm{Tr}_NH$ stationary. Since the only nonlinear parameter for
the trigonometric basis functions is $L$ we have
\begin{eqnarray}\label{Tr}
\frac{\mathrm{d}}{\mathrm{d}L}\mathrm{Tr}_NH=0.
\end{eqnarray}
Using Eq.~(\ref{even}) we can find $\mathrm{Tr}_NH$ as
\begin{eqnarray}
\mathrm{Tr}_NH=\frac{\pi^2}{12L^2}(4N^2-1)N+\left(\frac{L}{\pi}\right)^k\left[\sum_{n=1}^ND_{2n-1}+ND_0\right],
\end{eqnarray}
where $D_0=\pi^k/(k+1)$ and
\begin{eqnarray}
\sum_{n=1}^ND_{2n-1}&=&\sum_{i=0}^{\frac{k}{2}-1}\frac{\pi^{k-2i-2}k!}{(-4)^{i
+ 1}(k-2i-1)!}\Big[\left(4^{i+1}-1\right)\zeta(2i+2)
-\zeta(2i+2,N+1/2)\Big],
\end{eqnarray}
where $\zeta(s)$ and $\zeta(s,a)$ are the Riemann zeta function and
the generalized Riemann zeta function, respectively. After some
algebra, Eq.~(\ref{Tr}) results in the following optimal length and
$\alpha_{\mathrm{T}}$
\begin{eqnarray}
L_{\mathrm{T}}(N)&=&\pi\left(\frac{(4N^2-1)N/(6k)}{\displaystyle\frac{N\pi^k}{k+1}+\sum_{i=0}^{\frac{k}{2}-1}\frac{\pi^{k-2i-2}k!}{(-4)^{i
+
1}(k-2i-1)!}\Big[\left(4^{i+1}-1\right)\zeta(2i+2)-\zeta(2i+2,N+1/2)\Big]}\right)^{\frac{1}{k+2}},\nonumber\\
&\simeq&
\left(\frac{2(k+1)\pi^2}{3k}\right)^{\frac{1}{k+2}}N^{\frac{2}{k+2}},
\end{eqnarray}
\begin{eqnarray}
\alpha_{\mathrm{T}}(k)\simeq\frac{2(k+1)}{3k}.
\end{eqnarray}
For $k=2$ we have $L_{\mathrm{T}}(N)\simeq\sqrt{\pi N}$ which agrees
well with $L_S$ and $L_{\mathrm{op}}$. However, the accuracy of this
method reduces for $k>2$. In comparison with $L_{\mathrm{op}}$, we
should mention that for $k<10$, $L_{\mathrm{op}}$ is more accurate
as it is apparent from Fig.~\ref{fig2} and l.h.s of
Fig.~\ref{fig3}.\footnote{Note that for $k=2$, we have
$L_{\mathrm{op}}=L_S$, and for $k=4$, $L_{\mathrm{op}}$ nearly
coincides with $L_S$.} But for $k>10$, $L_{\mathrm{T}}$ is more
closer to $L_S$ (r.h.s of Fig.~\ref{fig3}) and similar to $L_S$
tends to one as $k$ goes to infinity. The accuracy of the energy
spectrum of the anharmonic oscillators for $k=\{2,4,6,8\}$ and
$N=10$ is reported in Table \ref{tab2}. As the table shows, the
optimal length obtained using $L_S$ or $L_{\mathrm{op}}$ formulas
gives more accurate results in comparison with $L_{\mathrm{T}}$
formula for a given $N$. In Fig.~\ref{fig5}, we depicted the
coefficient of proportionality $\alpha(k)$ for the proposed schemes.
Since $\alpha_{\mathrm{op}}$ grows exponentially, it cannot be used
efficiently for large $k$. On the other hand, $\alpha_S$,
$\alpha_{\mathrm{op}}^{(1)}$, and $\alpha_{\mathrm{op}}^{(2)}$ go to
$\pi^2/4$ and $\alpha_{\mathrm{T}}$ goes to $2/3$ at this limit.

\begin{table}
\begin{center}
\begin{tabular}{clc}
  \multicolumn{3}{c}{$k=4$} \\
  \hline
 $n$& $L$  &$\epsilon_n$
 \\\hline
0   &  $L_{S}$          &  1.53$\times10^{-11}$            \\
    &  $L_{\mathrm{op}}$&  1.65$\times10^{-11}$             \\
    &  $L_{\mathrm{T}}$ &  4.51$\times10^{-9}$             \\\\
2   &  $L_{S}$          &  1.88$\times10^{-10}$             \\
    &  $L_{\mathrm{op}}$&  2.10$\times10^{-10}$               \\
    &  $L_{\mathrm{T}}$ &  5.96$\times10^{-8}$              \\\\
4   &  $L_{S}$          &  6.05$\times10^{-9}$              \\
    &  $L_{\mathrm{op}}$&  6.59$\times10^{-9}$              \\
    &  $L_{\mathrm{T}}$ &  9.92$\times10^{-7}$              \\\\
6   &  $L_{S}$          &  2.12$\times10^{-7}$             \\
    &  $L_{\mathrm{op}}$&  2.08$\times10^{-7}$            \\
    &  $L_{\mathrm{T}}$ &  1.25$\times10^{-5}$              \\\hline
\end{tabular}
\quad\quad
\begin{tabular}{clc}
  \multicolumn{3}{c}{$k=6$} \\
  \hline
 $n$& $L$  &$\epsilon_n$
 \\\hline
0   &  $L_{S}$          &  1.57$\times10^{-9}$             \\
    &  $L_{\mathrm{op}}$&  1.14$\times10^{-10}$             \\
    &  $L_{\mathrm{T}}$ &  5.00$\times10^{-7}$              \\\\
2   &  $L_{S}$          &  9.03$\times10^{-9}$             \\
    &  $L_{\mathrm{op}}$&  2.48$\times10^{-9}$             \\
    &  $L_{\mathrm{T}}$ &  2.15$\times10^{-6}$              \\\\
4   &  $L_{S}$          &  8.07$\times10^{-8}$              \\
    &  $L_{\mathrm{op}}$&  8.93$\times10^{-8}$              \\
    &  $L_{\mathrm{T}}$ &  1.36$\times10^{-5}$              \\\\
6   &  $L_{S}$          &  4.88$\times10^{-7}$              \\
    &  $L_{\mathrm{op}}$&  1.58$\times10^{-6}$              \\
    &  $L_{\mathrm{T}}$ &  7.91$\times10^{-5}$              \\\hline
\end{tabular}
\quad\quad
\begin{tabular}{clc}
  \multicolumn{3}{c}{$k=8$} \\
  \hline
 $n$& $L$  &$\epsilon_n$
 \\\hline
0   &  $L_{S}$          & 4.10$\times10^{-8}$              \\
    &  $L_{\mathrm{op}}$& 7.36$\times10^{-8}$            \\
    &  $L_{\mathrm{T}}$ & 6.85$\times10^{-6}$              \\\\
2   &  $L_{S}$          & 1.30$\times10^{-7}$              \\
    &  $L_{\mathrm{op}}$& 1.91$\times10^{-7}$              \\
    &  $L_{\mathrm{T}}$ & 1.75$\times10^{-5}$              \\\\
4   &  $L_{S}$          & 7.04$\times10^{-7}$              \\
    &  $L_{\mathrm{op}}$& 5.70$\times10^{-7}$              \\
    &  $L_{\mathrm{T}}$ & 6.46$\times10^{-5}$              \\\\
6   &  $L_{S}$          & 4.03$\times10^{-6}$              \\
    &  $L_{\mathrm{op}}$& 8.73$\times10^{-7}$              \\
    &  $L_{\mathrm{T}}$ & 2.40$\times10^{-4}$              \\\hline
\end{tabular}
\end{center} \caption{The relative errors of various optimized schemes for $N=10$ basis functions.}\label{tab2}
\end{table}

\section{The harmonic oscillator basis expansion}\label{sec6}
\begin{figure}
\begin{center}
\includegraphics[width=8cm]{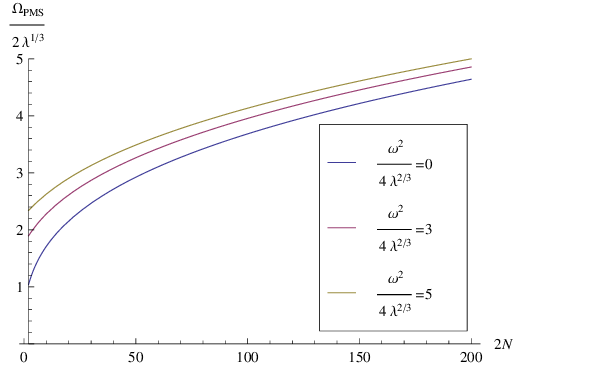}
\caption{\label{fig8}The optimal frequency parameter versus $2N$
obtained via the principle of minimal sensitivity for
$\omega^2=\{0,12\lambda^{2/3},20\lambda^{2/3}\}$.}
\end{center}
\end{figure}

We can also find the energy spectrum of the anharmonic oscillators
using the basis of the harmonic oscillator eigenfunctions
\begin{eqnarray}
\phi_n(x)=\left[\frac{\Omega}{\sqrt{\pi}2^nn!}\right]^{1/2}H_n(\Omega
x)\,e^{-\Omega^2x^2/2},
\end{eqnarray}
where $H_n(\Omega x)$ are Hermite polynomials and the frequency
$\Omega$ is the variational parameter. This frequency can be fixed
by the principle of minimal sensitivity, requiring the dependence on
the variational parameter be as weak as possible \cite{7}. For the
Hamiltonian
\begin{eqnarray}
H=\frac{1}{2}\left[-\frac{\mathrm{d}^2}{\mathrm{d}x^2}+\omega^2
x^2\right]+\lambda x^4,
\end{eqnarray}
the application of the PMS to the sum of $N$ even and $N$ odd basis
functions gives \cite{Okopinska}
\begin{eqnarray}
\Omega_{\mathrm{PMS}}^3-\omega^2\Omega_{\mathrm{PMS}}=8\lambda\left(N+\frac{1}{8N}\right).\label{pms}
\end{eqnarray}
It is also possible to find a similar relation for the optimal
frequency $\Omega_{\mathrm{op}}$ using the prescription presented in
Sec.~\ref{sec3}. Here we have two potentials: one physical
$\frac{1}{2}\omega^2 x^2+\lambda x^4$ and one unphysical
$\frac{1}{2}\Omega^2 x^2$ which we use the eigenfunctions of the
latter to approximate the solution of the former. The intersection
points of these potentials are given by
$x_{\mathrm{int}}=\pm\sqrt{\frac{\Omega^2-\omega^2}{2\lambda}}$. So
the potentials take the following value at these points:
\begin{eqnarray}
V(x_{\mathrm{int}})=\frac{\Omega^2}{4\lambda}\left(\Omega^2-\omega^2\right).\label{v1}
\end{eqnarray}
On the other hand, for  $N$ even and $N$ odd basis functions, the
maximal quantum number is $2N-1$ and therefore the maximal energy
reads
\begin{eqnarray}
E_{\mathrm{max}}=\Omega\left(2N-\frac{1}{2}\right).\label{v2}
\end{eqnarray}
Now since the basis functions with the energy larger than
$V(x_{\mathrm{int}})$ would have no useful contribution to the
sought-after solutions, by equating Eq.~(\ref{v1}) and
Eq.~(\ref{v2}) we obtain the relation for the optimal frequency as
\begin{eqnarray}
\Omega_{\mathrm{op}}^3-\omega^2\Omega_{\mathrm{op}}=8\lambda\left(N-\frac{1}{4}\right),
\end{eqnarray}
which agrees well with Eq.~(\ref{pms}) especially for large $N$. At
this limit, we have
$\Omega_{\mathrm{op}}\approx\Omega_{\mathrm{PSM}}\approx 2(\lambda
N)^{1/3}$. Also, the $N$ dependence of the optimal frequency
coincides with the $N$ dependence of the optimal length in the
trigonometric expansion for $k=4$, i.e., $\Omega_{\mathrm{op}}\sim
N^{1/3}\sim L_{\mathrm{op}}$. Note that, the accuracy of
$\Omega_{\mathrm{op}}$ is of the order of $\Omega_{\mathrm{PSM}}$
even for small $N$. In Fig.~\ref{fig8}, we have depicted
$\frac{\Omega_{\mathrm{PSM}}}{2\lambda^{1/3}}$ in terms of the total
number of the basis functions $2N$ for
$\omega^2=\{0,12\lambda^{2/3},20\lambda^{2/3}\}$.

\section{Conclusions}\label{sec7}
In this paper, we presented several optimal lengths for accurate
calculation of the eigenvalues and eigenfunctions of the anharmonic
oscillators using trigonometric basis functions obeying Dirichlet
boundary condition and the harmonic oscillator eigenfunctions. We
indicated that the value of the potential at the optimal length
should be proportional to the maximum energy of the used basis
functions. Since both $V(L_{\mathrm{op}})\gg \varepsilon_N$ and
$V(L_{\mathrm{op}})\ll \varepsilon_N$ are two sources of error, we
demanded that the proportionality coefficient $\alpha(k)$ to be of
the order of one. For the trigonometric basis set, we suggested some
ansatz for $\alpha(k)$ with the condition $\alpha(2)=1$ and found
that $L_{\mathrm{op}}^{(2)}$ gives the most accurate results. By
defining $L_S$ we showed that it can be also used as an accurate
optimal length. Indeed, it is shown that the optimal number of the
mesh points (for fixed $h$) in the Schwartz's scheme, where the
reference function vanishes there nearly coincides with the optimal
number of the basis functions (for fixed $L$) which at most have the
same number of nodes. Moreover, $\alpha_S$ is of the order of unity
for all $k$. An alternative proposal is using the trace of the
Hamiltonian as an only physical quantity before diagonalization and
to make it stationary at the optimal value $L_{\mathrm{T}}$. We
showed that this optimal length has the correct asymptotic value for
large $k$ and can be used as a good approximation for $k>10$. For
all proposals except $L_{\mathrm{op}}$ the proportionality
coefficient $\alpha(k)$ remains of order of one for all $k$ and they
have the correct asymptotic value, i.e.,
$\lim_{k\rightarrow\infty}L(N)=1$. We indicated that these proposals
can be also used for the multidimensional problems, and for large
$N$, for the periodic boundary condition and the polynomial
potentials. The following schematic diagram shows the relative
efficiency of the proposed optimal lengths
\begin{eqnarray}
\begin{array}{ccc}
L_{\mathrm{T}}  &  \xrightarrow{k<10}     & L_{\mathrm{op}}\\\\
L_{\mathrm{op}} &  \xrightarrow[k>10]{}   & L_{\mathrm{T}}
\end{array}
\longrightarrow \,\,\,\, L_{\mathrm{op}}^{(1)}\longrightarrow
\,\,\,\, L_S \,\,\longrightarrow \,\,\,\, L_{\mathrm{op}}^{(2)},
\end{eqnarray}
where the right arrow indicates the direction of the increasing of
the accuracy. For the harmonic oscillator basis functions, we showed
that the resulting optimal frequency agrees with the one obtained
using the principle of the minimal sensitivity.

\section*{Acknowledgement}
I would like to thank the referee for giving such constructive
comments and suggestions which considerably improved the quality of
the paper.

\end{document}